\documentclass[seceq,mbf,wrapft]{ptptex}

\usepackage{wrapft}

\def\beqra{\begin{eqnarray}}
\def\eeqra{\end{eqnarray}}
\def\beqast{\begin{eqnarray*}}
\def\eeqast{\end{eqnarray*}}
\def\be{\begin{enumerate}}
\def\ee{\end{enumerate}}
\def\lag{\langle}
\def\rag{\rangle}

\def\beq{\begin{equation}}
\def\eeq{\end{equation}}
\def\be*{\begin{equation*}}
\def\ee*{\end{equation*}}
\def\haf{\frac{1}{2}}

\def\nd{\noindent}

\notypesetlogo  

\title{
On the Weak Decay of Composite System\\ based on Majorana Partners
}

\author{
Masakazu {\sc Aoki} and Kazuo {\sc Koike}
}

\inst{
Department of Natural Science,~Kagawa University

Takamatsu 7608522, Japan
\\
}

\abst{
 In a composite model of leptons and quarks, composite particles with 
heavy Majorana particles appear provided that the V-rishon is regarded as a Majorana
particle. In this paper, the decay mode of such ``Majorana partners" is investigated.
It is shown that the life time of Majorana partners are fairly long.
Our model suggests that a new mechanism of proton decay is possible, where the X and Y
bosons in GUTs are not necessary.
}

\begin{document}

\maketitle

\section{Introduction}

 In the progress of high-energy physics, the so-called energy frontier approaches
to TeV region, where discoveries of new phenomena and/or new particles such as
the super-partners are expected.
In such situation, we have proposed in a few years ago a model of leptons and 
quarks on the basis of possible Majorana-partner in rishon model\cite{rf:schem}.  
Our model is,
however, in the stage of formal proposal and characteristics concerning to the 
decay properties etc are not yet made clear. When a new particle is observed, 
the decay properties will play an important role in the decision of its nature.
This paper is concerning to this problem. 

It should be noted that the treatment of composite system is generally complicated
compared with the simple system based on field theory such as SUSY.
However, the weak interaction of composite system had been formulated in an
appropriate way containing semi-empirical approach in the development of 
particle physics in the stages when little is known concerning to the nature of 
constructive force of composite system. 
In this paper, we will in first summarize the essence of our model. In section 3, 
quick review the treatments of weak interaction of composite system is given.  
Then, weak decay properties of our composite system are investigated.
The discussion is given in the final section.

\section{Composite model of leptons and quarks based on Majorana partner}

The essence of our model is summarized as follows.
In the rishon model,the leptons and quarks are given as\cite{rf:rishon} 

\beq
        { u =  TTV,~~
          d =  \tilde{V}\tilde{V}\tilde{T},~~}
          {\nu}_e = {VVV,~~
          e^- = \tilde{T}\tilde{T}\tilde{T}}
          \label{eq:Rishon}
\eeq
where the $V$ rishon is purely electrically neutral, then
it is possible to introduce the Majorana property on $V$.
It should be noted that we have used symbolically the notation $\tilde{V}$
as the anti-particle of $V$. Though the particle and anti-particle are identical
in the Majorana neutrino theory, they are distinguished historically and 
phenomenologically through the difference of their interactions by that of helicity. 
In fact in the Majorana neutrino theory, the difference between ``neutrino" and 
``anti-neutrino" is reduced to the difference of helicity state\cite{rf:Bilen}. 
In this paper, we use the symbol $\tilde{V}$ in this meaning.

For the $V$ rishon, let
us consider the Dirac-Majorana (D-M ) mass term 
\cite{rf:Bilen}\tocite{rf:Wol} in the simplest case of
one generation. 
We have\footnote{In a classification scheme based on the hyper-color and
color group
$ SU_3(H) \times SU_3(C) $,
the singlet property of the Majorana mass term may be destroyed if the rishon
belongs to the fundamental representation. The {\bf 8} representation 
seems to avoid this difficulty.
The details of the resolution of this problem will be discussed elsewhere.}

\begin{eqnarray}
{\cal{L}}^{\rm{D\mbox{-}M}}& = &
-\frac{1}{2} m_{L} \overline{(V_{L})^c} V_{L}
-m_{D}\bar{V}_{R}V_{L}
-\frac{1}{2} m_{R} \bar{V}_{R} (V_{R})^c~~+~~\rm{h.c.}
\label{eq:DM}
\end{eqnarray}
 From 
Eqs.~(\ref{eq:DM}) 
 we have
\beq
{\cal{L}}^{\rm{D-M}}=-\frac{1}{2} \sum_{\alpha=1}^{2}
 m_{{\alpha}}{\bar{\chi}}_{{\alpha}} \chi_{{\alpha}}
\label{eq:dia2}
\eeq
\noindent where
\begin{eqnarray}
V_{L}~~  =~~~{\cos{\theta}{\chi_{1L}}} & + & {\sin{\theta}\chi_{2L}}
\nonumber\\
(V_{R})^c  =  {-\sin{\theta}\chi_{1L}} & + & {\cos{\theta}\chi_{2L}}.
\label{eq:MIX}
\end{eqnarray}
\noindent Here $\chi_{1} $ and $\chi_{2} $ are fields of Majorana
 $V$ rishon with masses $ m_{s}~(a~~``small"~ mass)$, $m_{B}~(a~~``Big"~ mass) $, 
 respectively.
Assuming now that

\beq
m_{L}=0,~~~ m_{D}\simeq{ m_{F}},~~~m_{R}\gg{ m_{F}} 
\label{eq:mass-sb1}
\eeq

\noindent and

\beq
m_{s}\simeq\frac{m_{D}^2}{m_{R}},~~m_{B}\simeq{m_{R}}
{}~~\theta\simeq{\frac{m_{D}}{m_{R}}}
\label{eq:mass-sb2}
\eeq

\noindent where $m_{F} $ is a typical mass of the rishon,
particles with definite masses are distinguished as a very light Majorana
$V$ rishon with mass
$m_{s} \ll m_{F} $ and a very  heavy Majorana particle with mass
$m_{B}\simeq m_{R} $. The current $V$ rishon field, $V_{L} $, nearly
coincides with
 $\chi_{1L} $, and $ \chi_{2}\simeq {V_{R}~+~(V_{R})^c} $, 
because $\theta$ is extremely small.
It should be noted that $V_{L}$ and $(V_{R})^c$ represent the states corresponding
to that concerning the possible weak interaction, 
while  $\chi_{1} $ and $ \chi_{2}$ are the mass
eigenstates.
Hereafter, we use the notation
\beq
          V_s \equiv \chi_{1}~~~({\rm mass}~ m_s)~~~~~\\ 
          V_B \equiv \chi_{2}~~~({\rm mass}~ m_B) 
\label{eq:VsVB}
\eeq

In our scheme, the $V$ rishon $V_s$ is a Majorana
particle with mass much smaller than those of the other fermions.
The predictions for the $V$ rishon mass depend on the value of the
$m_{R}$ mass.

In the Dirac-Majorana mass scheme,  the rishons $V_s$ and $V_B$  appear.
In our scheme, the $V$ rishon $V_s$ is a Majorana
particle with mass much smaller than those of the other fermions.
The predictions for the $V$ rishon mass depend on the value of the
$m_{R}$ mass.

The existence of $V_B$ in Eq.~(\ref{eq:VsVB}) implies the new 
generation structure with additional particles shown 
in Table~\ref{table:1}, where $TTV_s$, 
$\tilde{T}\tilde{V}_s\tilde{V}_s$, $V_sV_sV_s$
and $\tilde{T}\tilde{T}\tilde{T}$ represent the observed quarks and leptons,
$u$, $d$, $\nu$ and $e^-$, respectively.
In addition, new heavy quarks $TTV_B$, $~\tilde{T}\tilde{V}_s\tilde{V}_B$ and
$\tilde{T}\tilde{V}_B\tilde{V}_B$ and heavy neutrinos 
$V_sV_sV_B$, $V_sV_BV_B$ and$V_BV_BV_B$ appear. They are characterized by their
configurations containing the heavy $V$ rishon $V_B$.
The row corresponding to $\nu$ lists the mass eigenstates of neutrinos belonging 
to the first
generation, and $\nu_e$ the state concerning the weak interaction, 
which is realized as a superposition of mass eigenstates.
\begin{table}
\begin{center}
\caption{Configuration of leptons and quarks constituting a generation,
where the row corresponding to $\nu$ lists its mass eigenstates belonging to the 
first generation.
}
\label{table:1}
\begin{tabular}{cccccccc} \hline \hline
flavor   &standard      & $B$            & $BB$          & $BBB$ \\ \hline
$u$      &$TTV_s$       & $TTV_B$ \\
$d$      &$~\tilde{T}\tilde{V}_s\tilde{V}_s$   & $~\tilde{T}\tilde{V}_s\tilde{V}_B$
 & $\tilde{T}\tilde{V}_B\tilde{V}_B$\\
$\nu$  &$~~V_sV_sV_s$   & $~~V_sV_sV_B$  & $~V_sV_BV_B$ & $~~V_BV_BV_B$\\
$~~e^-$    &$\tilde{T}\tilde{T}\tilde{T}$\\ \hline   
\end{tabular}
\end{center}
\end{table}

It should be emphasized that the new scheme satisfies the anomaly-free condition
at the level of leptons and quarks;
\footnote{ In this paper, we have assumed that the gauge structure appears
at the level of leptons and quarks, as a result of the dynamics of the sub-system,
and the anomaly-free condition is formally realized at this level. In this sense,
the meaning of this condition in the sub-system seems to be rather ambiguous.}
that is, the condition
$\sum{Q} = 0$
is satisfied.

It should be noted that
in the seesaw mechanism, if the mass of $V_B$ is of the order of the Planck
mass, the generation structure will not be affected in practice. The most 
interesting
case is in which the mass of $V_B$ is not extremely large,
for example, on the order of possible super-partners.
In such a case, heavy quarks and neutrinos appear at the level of leptons
and quarks.
In the special case in which the two Majorana particles have exactly the same
mass, $m_{V_s} = -m_{V_B}$, they will behave as if a single Dirac particle did,
which is often called a ``pseudo Dirac particle".

\section{Electroweak interaction in composite system}

What form does our compositeness appear in electroweak interaction?
We will examine the case when the compositeness
of leptons and quarks are extremely suppressed and leptons and quarks themselves 
behave as if elementary particles in the realization of electroweak structure.

\subsection{Electroweak interaction in quarks and nucleons}

In order to investigate the electroweak interaction of composite system, we will firstly compare the case of quark doublet ${u \choose d}$ with that of hadron doublet ${p \choose n}$, where $p = (uud)$ and $n= (udd)$. 

Under $SU(2)_L \times U(1)$, the weak doublet of quarks u and d is given as

\beq
q_{\small L}={u \choose d}_L
\label{eq:quark_db}
\end{equation}

\noindent
where the electric charge is  $Q = I_3 + Y$, and $Y = 1/6$.
After the standard procedure,the weak charged current is given as

\beq
J_{\mu} = \bar{q}\gamma_{\mu}(1 + \gamma_5){\frac{1}{2}} (\tau_1 + i\tau_2)~q~+ h.c.~~\\
\label{eq:q-current}
\eeq

\noindent
Explicit representation of Eq.~(\ref{eq:q-current}) is 
\beq
J_{\mu} = \bar{u}\gamma_{\mu}(1 + \gamma_5)~d~+ h.c.~~\\
\label{eq:q-current2}
\eeq

\nd
In the charged weak current of hadrons, it should be noted that proton and neutron are composite system of quarks, $p~=~(uud)$ and $n~=~(udd)$. The elementary process is caused by quarks, then the basic structure of weak current of hadrons is the same as that of quarks. However, the weak current of hadrons are affected by the compositeness.
\beq
J_{\mu} = \bar{p}\gamma_{\mu}(C_V + \gamma_5 C_A)d,~+ h.c.~~\\
\label{eq:h-current}
\eeq

\nd
Because of the conservation of vector current, $C_V$ is not renormalized, while the axial vector current is only partially conserved, 
\footnote{
The CVC and PCAC can be easily shown by making use of model Lagrangian.
The Lagrangian of u and d quarks is given as 
\begin{eqnarray*}
{\cal{L}}& = &
-\bar{q}(-i\gamma^{\alpha}\partial_{\alpha} + m_q)q  + {\cal{L}}_I
\label{eq:quark}
\end{eqnarray*}

\noindent
where ${\cal{L}}_I$ is the interaction term where derivative is generally not
contained. 
\nd
For the infinitesimal transformation
\be*
q \longrightarrow (1 + {i\over 2} {\tau}_{k} u_k)q\\
\ee*
\be*
J^{\alpha}_k = \bar{q}\gamma^{\alpha}{\haf}{\tau}_{k}q,~~~\\
\label{eq:Vcurrent}
\ee*
\be*
{\partial}_{\alpha}J^{\alpha}_k = 0. 
\ee*
For the axial-vector current, the PCAC relation is obtained, 
\be*
q \longrightarrow (1 + {i\over2} {\gamma}^5{\tau}_{k} u_k)q\\
\ee*
\be*
J^{5\alpha}_k = \bar{q}\gamma^{\alpha}{\gamma}^5{\haf}{\tau}_{k}q,\\
\label{eq:Acurrent}
\ee*
\be*
{\partial}_{\alpha}J^{5\alpha}_k = im_q\bar{q}{\gamma}^5{\tau}_{k}q. 
\label{eq:PCAC}
\ee*


}
then $C_A$ is renormalized. 

It should be noted that the same electroweak structure of proton and neutron is obtained provided that the weak doublet 

\beq
N_{\small L}={p \choose n}_L
\label{eq:neucleon_db}
\end{equation}

\noindent
is taken, where the electric charge is  $Q = I_3 + Y$, and $Y = 1/2$.
In this case, the weak charged current is given as

\beq
J_{\mu} = \bar{p}\gamma_{\mu}(1 + \gamma_5)n,~+ h.c.~~\\
\label{eq:N-current}
\eeq

\nd
Eq.~(\ref{eq:N-current}) is same as Eq.~(\ref{eq:h-current}) except the factors $C_V$ and $C_A$.

That is, the effective electroweak structure with the same structure appears in the level of protons and neutrons\cite{rf:weak-o}. 

\subsection{Effective electroweak interaction in composite system}

Thus, we will consider the case in which the processes concerning the sub-structure  are extremely suppressed and composite leptons and quarks themselves behave as if elementary particles in the realization of electroweak structure.
In the original rishon model, the leptons and quarks in this model are given as\cite{rf:rishon} 
\beq
        { u =  TTV,~~
          d =  \tilde{V} \tilde{V} \tilde{T},~~}
          {\nu} = {VVV,~~
          e^- =   \tilde{T}\tilde{T} \tilde{T}}
          \label{eq:Rishon}
\eeq
where $\nu$ represents the neutrino belonging to mass eigenstate, and the state 
contributing to weak interaction  ${\nu}_e$ is a superposition of such states.
It should be noted that the $V$ rishon is purely electrically neutral, therefore
it is possible to introduce the Majorana interaction on $V$.

Though a few models are proposed to introduce the generation structure, in this paper, we will represent the generation structure byintroducing the generation suffix i for each rishon doublets.
Then, the leptons and quarks in our system are given as,

\beq
        { U_i =  T_iT_iV_i,~~
          D_i =  \tilde{V_i} \tilde{V_i} \tilde{T_i},~~}
          {\nu}_i = {V_iV_iV_i,~~
          l_i =   \tilde{T_i}\tilde{T_i} \tilde{T_i}}.
          \label{eq:Rishon_i}
\eeq

\noindent
In the case when the realization of electroweak structure is extremely suppressed by some reason, the relevant electroweak structure will appear only for composite system in Eq.~(\ref{eq:Rishon_i}).
The Lagrangian density in our such system is given as

\begin{eqnarray}
     {\cal L} = {\cal L_0} &-& \sum_{i=1}^3 ~\bar{L_i}\gamma_\mu(\partial_\mu - ig\mathbf{t}\cdot\mathbf{A_\mu} -ig^\prime YB_\mu)L_i 
- \sum_{i=1}^3 ~\sum_{\kappa=U,D}~\bar{R_i^\kappa}(\partial_\mu -ig^\prime YB_\mu){R_i^\kappa}
\nonumber\\
&-& \sum_{i,j=1}^3~(\bar{L_i}\phi R_j^D M_{ij}^D + h.c.)
- \sum_{i,j=1}^3~(\bar{L_i}(-\eta)\phi^* R_j^U M_{ij}^U + h.c.)
\nonumber\\
&+& {\cal L^\prime} (lepton~ term)
\label{eq:EW_composite}
\end{eqnarray}

\noindent 
where 

\beq
L_i = \frac {1 + \gamma_5}{2} {{U_i} \choose {D_i}},~~
R_i^U = \frac {1 - \gamma_5}{2} {U_i},~~
R_i^D = \frac {1 - \gamma_5}{2} {D_i}
\eeq

\noindent 
and
\begin{equation*}
\eta = \left( \begin{array}{cc}
 0 & I \\
-I & 0
\end{array} \right).
\label{eq:MAT_eta}
\end{equation*}

The scalar field $\phi$, the gauge field $\mathbf{A_\mu}$ and $B_\mu$ transform as a doublet, triplet and singlet, respectively under weak rotation $SU(2)_L$.

The effect of symmetry breaking is obtained by replacing
\beq
\phi^0 \to e^{i\theta} (v + \phi^0\prime)
\label{eq:Higgs}
\eeq

\noindent
where $v$ represent of the magnitude of vacuum expectation value of $\phi^0$ field,
\beq
\langle 0 \rvert \phi^0 \lvert =  e^{i\theta}v
\label{eq:Higgs_exp}
\eeq

Now, we must consider the diagonalization of the mass term
 ${\cal L}_{mass}$ obtained from Eqs.~(\ref{eq:EW_composite}) - (\ref{eq:Higgs_exp}),

\begin{eqnarray}
{\cal L}_{mass} = &-&
\nonumber 
\sum_{i,j=1}^3~\bar{D}_{Li}(e^{i\theta}vM_{ij}^D)D_{Rj} + h.c.)\\
&-& \sum_{i,j=1}^3~\bar{U}_{Li}(e^{-i\theta}vM_{ij}^U)U_{Rj} + h.c.)
\label{eq:EW_mass}
\end{eqnarray}

\noindent
Performing an appropriate unitary transformation,
\beq
{\cal L}_{mass} = 
 \bar{D}_{L}M_D D_{R} + \bar{U}_{L}M_U U_{R} +  h.c.~  
\label{eq:Unitary}
\eeq

\noindent
the mass term Eq.~(\ref{eq:Unitary}) can be diagonalized. The result  is  given 
as,
\beq
   U_i =\sum_{l} V_{il}^{(U)}U_l~ ,~   D_i =\sum_{k} V_{ik}^{(D)}D_k
\label{eq:Mass_uni}
\eeq

\noindent
 The mass matrices in Eq.~(\ref{eq:Mass_uni}) are given as,
 
\beq
    M_D = ( V^{(D)\dagger}(e^{i\theta}vM_{ij}^D)V^{(D)}) ~~ 
    M_U = ( V^{(U)\dagger}(e^{-i\theta}vM_{ij}^U)V^{(U)}) 
\label{eq:Mass_Matrix2}
\eeq

\noindent
where the prime mark of diagonalized fields  is  abbreviated  for 
simplicity of description. It should be noted that the CKM matrix is given as,
\beq
     K = V^{(U)}V^{(D)\dagger}   
\label{eq:CKM}
\eeq

\nd
It should be noted that our formulation of this section is parallel with the standard procedure. That is, though the fundamental system is composite, the composite
systems behave partially as if they are elementary particles.         
In what form, does rare process appear on the basis of the sub-structure ?

\eject

\section{Pre-electroweak structure for sub-constituents}

\begin{wraptable}{l}{\halftext}
\caption{Leptons and quarks belonging to the first generation.
}
\label{table:2}
\begin{center}
\begin{tabular}{cccccc} \hline \hline
&~$Q$        &flavor    & $B1$            & $B2$           & $B3$ \\ \hline
&$~2/3$    &$u$       & $u_{B_1}$ \\ 
&$-1/3$    &$d$       & $d_{B_1}$     & $d_{B_2}$ \\  
&$~0$      &~$\nu  $  & $~\nu_{B_1}$  & $\nu_{B_2}$  &$\nu_{B_3}$ \\
&$-1$     &~~$e^-$\\ \hline   
\end{tabular}
\end{center}
\end{wraptable}

In this section, we will treat the simplest case with one generation for simplicity.
Then, the weak-doublets in our system is given as,

\beq
        \Psi_L =  { T \choose  \tilde{V}}_L
          \label{eq:Rishon2}.
\eeq

In order to see the essential nature, we will in this section restrict to one 
generation case, because general mixing scheme in this case is complicated.

\noindent
The electroweak Lagrangian in our system is given as

\begin{eqnarray}
     {\cal L} =  &-& ~\bar{\Psi}_L \gamma_\mu(\partial_\mu - ig\mathbf{t}\cdot\mathbf{A^{(s)}_{\mu}} -ig^\prime YB^{(s)}_{\mu} )\Psi_L 
- \sum_{\kappa=T,V}~\bar{\Psi}_R^{\kappa }(\partial_\mu -ig^\prime YB^{(s)}_\mu){\Psi}_R^{\kappa }
\nonumber\\
&-& ~(\bar{\Psi}_L \phi \bar{\Psi}_R^{V } M^{(V)} + h.c.)
- (\bar{\Psi}_L(-\eta)\phi^* \bar{\Psi}_R^{T} M^{(T)} + h.c.)
\label{eq:EW_Sub}
\end{eqnarray}

\noindent
where $\mathbf{A}^{(s)}_\mu $ and $B^{(s)}_{\mu}$ are the gauge fields concerning to sub-system.
The electric charge of component of doublet Eq.~(\ref{eq:Rishon2}) is given as
\beq
Q = I_3  +  Y
\label{eq:charge_sub}
\eeq

\noindent
in the unit of e/3, and $Y = -1/2$.
After symmetry breaking, the sub-gauge bosons $W^{(S)\pm }_\mu$,  $Z^{(S)}_\mu$ and $A^{(s)}_\mu$ bosons  appear, where they are characterized by the charge unit e/3.
The ordinary electroweak gauge bosons $W_{\mu}$, $Z_{\mu}$ and $A_{\mu}$ will be formally expressed by making use of the sub-gauge bosons $W^{(s)}_{\mu}$, $Z^{(s)}_{\mu}$ and $A^{(s)}_{\mu}$ as,

\begin{eqnarray}
W_{\mu} &=& (W^{(s)}W^{(s)}W^{(s)})_{\mu}~\nonumber\\
Z_{\mu} &=& (Z^{(s)}Z^{(s)}Z^{(s)})_{\mu}~\nonumber\\
A_{\mu} &=& (A^{(s)}A^{(s)}A^{(s)})_{\mu}
\label{eq:effective_EW}
\end{eqnarray}

\noindent 
It should be noted that Eq.~(\ref{eq:effective_EW}) is a symbolical representation of
composite system based on yet to be known mechanism.
These sub-gauge bosons should be almost confined in the ordinary gauge bosons belonging  the level of leptons and quarks.
The effective electroweak structure in the level of leptons and quarks will be given by making use of Eq.~(\ref{eq:effective_EW}).

In order to investigate the properties of composite system given by 
Eq.~(\ref{eq:effective_EW}), let us examine the familiar $\beta$-decay process
of d quark,

\beq
d + {\nu}_e    \to  u + e^- 
\label{eq:beta}
\eeq

\noindent
In the picture of fundamental constituents, Eq.~(\ref{eq:beta}) is rewritten as,

\beq
(\tilde{V}\tilde{V}\tilde{T}) + (VVV) \to (TTV) + (\tilde{T}\tilde{T}\tilde{T})
\label{eq:beta-rishon}
\eeq

\noindent
Thus, we see that the fundamental interaction in our process is

\beq
{\cal L} = g_{w}^{(s)} \bar{T}\gamma^\mu (1 + \gamma_5 ) \tilde{V} W_\mu^{(s)}
\label{eq:fundamental_int}
\eeq

\noindent
In our scheme, the elementary process such as

\beq
\tilde{V} + V \to T + \tilde{T}
\label{eq:elementary_process}
\eeq

\noindent
is expected to be extremely suppressed as if quarks are almost completely
confined, while the process such as Eq.(\ref{eq:beta}) or equivalently  
Eq.(\ref{eq:elementary_process}) are enhanced similar to that the 3-body system
of quarks are familiar ones.

Thus, on the basis of such normalization assumption, we will investigate
possible decay mode of heavy particles with Majorana constituent $V_B$.

\section{ Weak decay of composite system with Majorana partners } 

It should be noted that the Majorana constituents in our model,
$V_s$ and $V_B$ are the mass eigenstates, which are represented as
superposition of the eigenstates of weak interaction, $V_{L}$ and $V_R^{c}$,

\noindent 
\begin{eqnarray}
V_{sL}~~ & = & {\cos{\theta}}V_{L}  -  {\sin{\theta}}V_R^{c}
\nonumber\\
V_{\tiny BL}~~ & = & {\sin{\theta}}V_{L}  +  {\cos{\theta}}V_R^{c}.
\label{eq:MIX_inv}
\end{eqnarray}

It should be emphasized that $V_R^{c}$ in Eq.(\ref{eq:MIX_inv}) does not
appear in ordinary weak interaction Lagrangian, and called as ``sterile'' 
particle.\cite{rf:Pontecorvo}

In order to investigate the $\beta$-decay process of our composite system,
let us compare the decay processes of $d$ and $d_{B_1}$.

The most familiar $\beta$decay process of neutron is known to be caused by
the $d \to u$ transition.
\noindent
By making use of Eq.(\ref{eq:MIX_inv}), the fundamental transition behind 
$d \to u$ transition as composite system
is symbolically represented as

\begin{eqnarray}
\noindent
(T\to &\tilde{V}&, T \to \tilde{V},V \to \tilde{T}) 
\nonumber\\
&=&(\bar{\tilde{V}}_L\gamma_\mu(1 + \gamma_5 ) T~cos{\theta},
\bar{\tilde{V}}_L\gamma_\mu(1 + \gamma_5 ) T~cos{\theta},
\bar{\tilde{T}}\gamma_\mu(1 + \gamma_5 ) V~cos{\theta}).
\label{eq:d to u}
\end{eqnarray}

\noindent
It should be noted that Eq.~(\ref{eq:d to u}) is physically equivalent with
the leptons and quarks level interaction

\beq
{\cal L} = g \bar{u}\gamma^\mu (1 + \gamma_5 ) d~W_\mu
\label{eq:d to u ql-level}
\eeq

\noindent
then the strength of the interaction given by Eq.~(\ref{eq:d to u}) should be
``normalized" to that of Eq.~(\ref{eq:d to u ql-level}) in a similar concept
as renormalization.

The transition behind $d_{B_1} \to u$ transition of composite system with
heavy constituent $V_B$ is also represented as

\begin{eqnarray}
\noindent
(T\to &\tilde{V}&, T \to \tilde{V_B},V \to \tilde{T} ) 
\nonumber\\
&=&(\bar{\tilde{V}}_L\gamma_\mu(1 + \gamma_5 ) T~ cos{\theta},
\bar{\tilde{V}}_L\gamma_\mu(1 + \gamma_5 ) T~ sin{\theta},
\bar{\tilde{T}}\gamma_\mu(1 + \gamma_5 ) V~ cos{\theta}).
\label{eq:dB1 to u}
\end{eqnarray}

\noindent
In our case, the mass of $V_B$ is expected to be very large. Then from 
Eq.~(\ref{eq:mass-sb2}),
$ cos{\theta} \sim 1$ and $ sin{\theta} \sim {\frac{m_{D}}{m_{R}}}$ . 
Thus, the strength of the process given by
Eq.~(\ref{eq:dB1 to u}) is distinguished by the factor $sin{\theta}$
compared with the process given by Eq.~(\ref{eq:d to u}).
In taking into account of the effect of unknown mechanism of composite system,
we will represent the effective suppression factor from the standard basic process 
given by Eq.~(\ref{eq:d to u}) as $f(sin{\theta})$, where the difference of phase 
volume effect due to the mass difference is not taking into account.
In a similar way, we see that the suppression factor of the processes
$u_{B_1} \to d$ and $d_{B_2} \to u$ is  $f(sin{\theta})$ and $f(sin^2{\theta})$,
respectively. The decay property of composite system with such constituents will 
be treated by making use of the method discussed in section 3.

In the same way, we see that the suppression factor of the typical heavy neutrino
decay
\beq
\nu_{B_1} \to e^+  + e^-  + \nu_e 
\label{eq:heavy nu}
\eeq

\noindent
is given by $f(sin{\theta})$, except by the difference due to the mass and phase
volume.


\section{ Possible proton-decay mechanism mediated\\ by sub-gauge boson 
$W^{(s)}$} 

It is interesting problem by which charge 1/3 or 4/3 is possible proton decay
mediated? It is known that the X boson with charge 4/3 mediates the proton decay
in GUTS such as $SU(5)$ model. In the previous papers, we have discussed that it is
possible to make a correspondence between GUTs picture and composite model.
\footnote{
It should be noted that model to explains
the proton-decay problem without difficulty. A typical decay mode is known as
\beq
P \to e^+  + {\pi}_0
\nonumber
\eeq
which is caused due to the $X$ gauge boson in GUTs.
In our model, this process is realized through the same process as in GUTs,
\beq
u + u \to X \to e^+ + \bar d
\nonumber
\eeq

\noindent
where we have regarded the 6 rishon system as $X$.
It should be noted that the gauge bosons are assumed to be 6-body systems of rishons
in the first work on the rishon model.\cite{rf:rishon}
The suppression of this process will be reduced to that of the rearrangement of 
rishons or, equivalently, the largeness of the $X$ boson mass.}

It should be emphasized, however, the new type of proton decay mediated by
the pre-gauge boson $W^{(s)}$ with charge 1/3 is possible in our model.
That is, the elementary process mediated by $W^{(s)}$ boson exchange,

\beq
V + \tilde{V} \to T + \tilde{T}
\label{eq:elementary}
\eeq

\noindent
the processes
\beq
(TTV) \to (TTT)~~,~~
(\tilde{V} \tilde{V} \tilde{T}) \to (\tilde{V} \tilde{T} \tilde{T})
\label{eq:induced}
\eeq

\noindent
which are represented as

\beq
u \to e^+ ~,~~
d \to \bar{u}
\label{eq:induced2}
\eeq

\noindent
are caused.

\noindent
Thus, the process 

\beq
(uud) \to (e^+ \bar{u}u)~
\label{eq:ProtonDecay0}
\eeq

\noindent
is induced, which means the proton decay process,

\beq
P \to e^+  +~ {\pi}_0
\label{eq:ProtonDecay}
\eeq

\noindent
is mediated by $W^{(s)}$ boson.
In our model the proton decay is caused without the X or Y gauge bosons appearing in GUTs.
It should be emphasized that in the proton decay of this type, the Majorana property
of V is essential. It is impossible in the case of Dirac type V where the 
anti-particle $\tilde{V}$ is essentially different from V.
In this sense, we will call our mechanism as ``Majorana decay of proton".







\section{Discussion}

  In this paper, we have investigated the properties of particles with 
the new heavy ``Majorana partners" $u_{B_1}$,~~$d_{B_1}$, ~~ $d_{B_2}$,
~~$\nu_{B_1}$,~~$\nu_{B_2}$ and $\nu_{B_3}$. These particles will be created
through pair creation by such familiar process as high-energy $PP$ collision. 
These particles are, however, fairly stable in the weak decay, and the suppression 
factor compared with the weak decay of ordinary particles is represented as
$f(sin{\theta})$ with $\theta\simeq{\frac{m_{F}}{m_{V_B}}}$, where ${m_{F}}$,
${m_{V_B}}$ is a typical mass of the rishon, heavy rishon, respectively.
The explicit form of function $f(sin{\theta})$ depends on the yet to be known dynamics of
sub-level, nevertheless $f(sin{\theta})$ dependence of decay process will suggest 
the appearance of Majorana partners when new particles are discovered in TeV or more
high-energy region.
It should be noted that our investigation is based on the standard model with (V-A)
current structure, where heavy Majorana particle $V_R^{c}$ appears as ``sterile" 
particle. In a vector-like theory\cite{rf:Vector-like} such as $\rm SO(10)$ model, 
right-hand current appears. In such case, new rare process of $V_R^{c}$ decay is 
possible.

It should be emphasized that our model is based on the extension of 
the knowledge for existence-form of matter found in Majorana 
neutrinos\cite{rf:Bilen,rf:GRSY,rf:SK-Atmospher}
to other fundamental particle such as the $V$ rishon as a universal property. 
Their possible existence should be made clear in the near future.  

Our work is based on the rishon model, which is still at a hypothetical level.
However, this model is based on the simple concept of the elementarity of electric 
charge, and in this sense this model is natural. In fact,
five types of electric charge appear in GUTs,\cite{rf:SU5} 
\beq
Q ~=~~0,~~~~1/3,~~~~2/3,~~~~1,~~~~4/3 
\eeq
where the 4/3 charge is carried by $X$ gauge boson, which is 
considered to be the particle responsible for proton decay.

Then, as far as the charge units are concerned, it seems that 
GUTs are too complicated to be the final form of the theory of elementary particles.

It should be noted that the standard formulation is not yet known in the 
rishon model, and, as mentioned above, it is probable that the rishon is beyond 
ordinary quantum field theory.\cite{rf:Maki} 
However, it seems to be meaningful to treat it in the framework of conventional 
field theory.
The formulation on sections 2 and 3 is along this treatment and it is assumed that  
the gauge structure of the Standard Model appears 
as a result of behavior of the level of leptons and quarks.

 It seems that there is another possible approach, in which, the pre-electroweak 
structure appears in the rishon level, and a certain kind of confinement of 
pre-gauge bosons realizes the well-known electroweak structure at the level 
of leptons and quarks. In section 4, we have step into such approach, however,
only a few important pictures are clarified by making use of some symbolical 
representations, because the fundamental dynamics of sub-system is not yet known.
Why are the sub-gauge bosons confined in $W_\mu, Z_\mu$ and $A_\mu$ bosons?
A possible approach to the ``enhancement" of these bosons is to introduce sub-colors
to such pre-gauge bosons.

We have shown that our model predict a new type of proton decay mechanism mediated
by the pre-gauge boson $W^{(s)}$ with electric charge 1/3. The extreme suppression
of this process will be explained by sub-color confinement. It should be emphasized
that the proton decay of this type is mediated by the boson with charge 1/3, in 
contrast to the decay in GUTs, where the proton decay is mediated by gauge bosons
X and Y. Especially, it should be noted that the charge of X boson is 4/3.
If we are fully in the model with sub-structure, GUTs is itself the effective theory
or phenomenological theory of composite system based on sub-system, where the 
fundamental electric charge is,
\beq
Q ~=~~0~~,~~1/3
\eeq

Finally, is there really a sub-level below the level of leptons and quarks? 
In our opinion, such a sub-level surely exists, and disclosing it will make it 
possible
to predict theoretically quantities such as the Higgs coupling constants,
the magnitudes of symmetry breaking
$\lag\phi\rag$, the mass spectrum and mixing parameters of all particles, etc.
We conclude this paper by emphasizing that the possible existence of 
Majorana partners is probable, together with that of the super-partners in SUSY.

\section*{Acknowledgments}
 This work was supported by Grant-in-Aid of Japanese Ministry of 
 Education, Science, and Culture (15540384). 



\end{document}